\documentclass[12pt]{iopart}

\usepackage{graphicx}% Include figure files
\usepackage{dcolumn}% Align table columns on decimal point
\usepackage{bm}% bold math

\begin{document}

\title[A Perfect Lens for Ballistic Electrons]{A Perfect Lens for Ballistic Electrons: A Kane's Model}

\author{I. Hrebikova$^1$, L. Jelinek$^1$, J. Voves$^2$ and J. D. Baena$^3$}

\address{$^1$ Department of Electromagnetic Field, Czech Technical University in Prague, Prague, Czech Republic}

\address{$^2$ Department of Microelectronics, Czech Technical University in Prague, Prague, Czech Republic}

\address{$^3$ Department of Physics, National University of Colombia, Bogota, Colombia}

\ead{hrebiiva@fel.cvut.cz}

\begin{abstract}
The analogy between electromagnetic waves and ballistic electrons within the Kane's model is developed and subsequently applied to a theoretical description of a quantum version of a metamaterial planar lens. Restrictions imposed on the perfect lens and the poor man's lens by available semiconductor band structures are discussed. A realistic implementation is proposed for the quantum poor man's lens, which uses specific properties of the HgTe compound. The properties of the lens are presented on the basis of a calculated transmission of oblique electrons through the lens structure.
\end{abstract}


\maketitle

\section{Introduction}
Analogies between light waves and electron waves have a long history, starting in the early years of quantum mechanics \cite{Slater-1928} and culminating in extensive review papers \cite{Henderson-1991}, \cite{Dragoman-1999} during the 1990s. At that time it was already well understood that there is an almost exact analogy between electromagnetic plane waves and quantum mechanical waves describing ballistic non-relativistic electrons. In fact, as present-day semiconductor techniques allow for precise monolayer growth, ballistic transport is routinely observed, and many semiconductor devices based on this analogy have been developed and experimentally tested. As representative examples we mention electrostatic lenses \cite{Spector-1990a}, \cite{Sivan-1990}, prisms \cite{Spector-1990b}, directional couplers \cite{Alamo-1989}, \cite{Thomas-1993}, filters \cite{Gaylord-1989} and circuit theory concepts \cite{Khondker-1988}. Unfortunately, developments in this field peaked well before the emergence of metamaterials, i.e. composite materials offering electromagnetic properties not found in natural substances, such as negative permittivity and permeability \cite{Marques}, \cite{Solymar}. 

Metamaterials have brought important new concepts into classical electromagnetism, e.g. the perfect lens \cite{Pendry-2000} and transformation optics \cite{Leonhardt-2006}, \cite{Pendry-2006}, while their quantum analogies have not received the attention that they deserve, with the exception of a few pioneering studies that will be briefly reviewed. In \cite{Kobayashi-2006}, a particularly simple form of the analogy of an electromagnetic plane wave and an electron wave is presented. The author then transfers the idea of complementary media into the electron domain using transmission matrix formalism, and proposes useing the complementary medium layer to improve the scanning tunneling microscopy of specific structures. Paper \cite{Dragoman-2007} uses the analogy presented in \cite{Kobayashi-2006} to explore the \mbox{I--V characteristics} and the traversal times of ballistic electrons propagating normally to the boundaries of the heterostructure analogous to the metamaterial perfect lens. The electron analogy of a perfect lens was further proposed in the form of a \mbox{p--n junction} on a graphene sheet \cite{Cheianov-2007}. In \cite{Zhang-2008} it was shown that spatial transformations leading to transformation optics and to metamaterial cloaking can also be used in a very similar manner used on the Schr{\"o}dinger equation. Later, in \cite{Jelinek-2011_b}, it was shown that ballistic electrons propagating in \mbox{the HgTe--CdTe} heterostructure can exhibit perfect tunneling, a phenomenon largely responsible for the unique properties of the perfect lens. Important papers \cite{Silveirinha-2012a}, \cite{Silveirinha-2012b} then showed that envelope approximation, commonly used for describing ballistic electrons in semiconductor heterostructures, is equivalent to the effective medium theory commonly used for describing of electromagnetic metamaterials. By means of this effective medium, a perfect lens made of graphene \cite{Silveirinha-2012a} was proposed. Lastly, the analogies mentioned above were used for a study of the cloaking of matter waves \cite{Fleury-2012}, \cite{Silveirinha-2012c}, \cite{Fleury-2013}.

In this paper, we will discuss the quantum analogy of the perfect lens \cite{Pendry-2000}. For this purpose, the analogy between plane waves and ballistic electrons presented in \cite{Jelinek-2011_b} will be extended to obliquely incident electrons. The quantum description of a semiconductor will be based on the 4-band Kane's model, the parameters of which are fixed by microscopic pseudopotential calculations. The demands on the semiconductor band structure for achieving perfect lensing are discussed, and finally we propose realistic devices using HgTe, a semiconductor with an inverted band structure.

%---------------------------------------------------------------------------------------------------
%---------------------------------------------------------------------------------------------------
\section{An Analogy Between Electron Waves and Light Waves}
Before getting to the actual lens design, a formal analogy between ballistic electrons and electromagnetic plane waves has to be established. The topology of a perfect lens \cite{Pendry-2000} is represented by a layered isotropic medium (stacking along the $z$-axis is assumed) in which oblique plane waves with the wavevector ${\bf{k}} = {k_z}{{\bf{z}}_0} + {k_{\mathrm{y}}}{{\bf{y}}_0}$ (${k_{\mathrm{y}}}$ being the transversal wavenumber) propagate. In the quantum domain, such a heterostructure is commonly described by the 4-band Kane's model \cite{Bastard}, where these bands correspond to conduction electrons (symmetry $\Gamma_6$), light and heavy holes (symmetry $\Gamma_8$) and split-off holes (symmetry $\Gamma_7$). Then using the approximations suggested in \cite{Bastard-1982}, which involves dropping the freespace terms for the $\Gamma_7$ and $\Gamma_8$ bands, the spin states of the conduction electrons remain degenerated even at oblique incidence, and are described by a scalar equation for the wavefunction $f_\mathrm{c}$
%%%%%%%%%%%%%%%%%%%%%%%%%%%%%%%%% Equation 1 %%%%%%%%%%%%%%%%%%%%%%%%%
\begin{eqnarray}
	\label{eq1}
		\left[ { - \frac{{{\hbar ^2}}}{{2m}}\frac{{{\partial ^2}}}{{\partial {z^2}}} + \frac{{{\hbar ^2}}}{{2m}}k_y^2 + {E_{{\Gamma _6}}} - E}	\right]{f_{\mathrm{c}}}\left( z \right) = 0,
\end{eqnarray}
%%%%%%%%%%%%%%%%%%%%%%%%%%%%%%%%%%%%%%%%%%%%%%%%%%%%%%%%%%%%%%%%%%%%%
where
%%%%%%%%%%%%%%%%%%%%%%%%%%%%%%%%% Equation 2 %%%%%%%%%%%%%%%%%%%%%%%%%
\begin{eqnarray}
	\label{eq2}
	\frac{1}{m} = \frac{{2{P^2}}}{3}\left( {\frac{2}{{E - {E_{{\Gamma _8}}}}} + \frac{1}{{E - {E_{{\Gamma _7}}}}}} \right)
\end{eqnarray}
%%%%%%%%%%%%%%%%%%%%%%%%%%%%%%%%%%%%%%%%%%%%%%%%%%%%%%%%%%%%%%%%%%%%%
is the inverse of the mass of the electron inside the material, $k_y$ is the transversal wavenumber and $E$ is the energy of the electron. $E_{\Gamma_6}$, $E_{\Gamma_7}$ and $E_{\Gamma_8}$ are the band edge energies of the $\Gamma_6$, $\Gamma_7$ and $\Gamma_8$ bands. Parameter $P$, known as the Kane's parameter, does not depend on energy, is well defined for commonly--used semiconductors, and can be obtained from experiments or, as in this paper, from microscopic calculations. It is also important to note that the validity of (\ref{eq1}), (\ref{eq2}) was recently confirmed by \cite{Silveirinha-2012b}, where it was derived by applying a homogenization technique. Following \cite{Kobayashi-2006}, the equation (\ref{eq1}) is rewritten in matrix form
%Schr{\"o}dinger 
%%%%%%%%%%%%%%%%%%%%%%%%%%%%%%%%% Equation 3 %%%%%%%%%%%%%%%%%%%%%%%%%
\begin{eqnarray}
	\label{eq3}
		\frac{\partial }{{\partial z}}\left[ {\begin{array}{*{20}{c}} {{f_{\mathrm{c}}}}\\
{\frac{{ - {\mathrm{i}}\hbar }}{m}\frac{{\partial {f_{\mathrm{c}}}}}{{\partial z}}}
\end{array}} \right] = \left[ {\begin{array}{*{20}{c}}
0&{{\mathrm{i}}\frac{m}{\hbar }}\\
{2{\mathrm{i}}\frac{{\left( {E - {E_{{\Gamma _6}}} - \frac{{{\hbar ^2}}}{{2m}}k_{\mathrm{y}}^2} \right)}}{\hbar }}&0
\end{array}} \right]\left[ {\begin{array}{*{20}{c}}
{{f_{\mathrm{c}}}}\\
{\frac{{ - {\mathrm{i}}\hbar }}{m}\frac{{\partial {f_{\mathrm{c}}}}}{{\partial z}}}
\end{array}} \right]
\end{eqnarray}
%%%%%%%%%%%%%%%%%%%%%%%%%%%%%%%%%%%%%%%%%%%%%%%%%%%%%%%%%%%%%%%%%%%%%
and supplemented by boundary conditions, namely \cite{Bastard}, by the continuity of $f_\mathrm{c}$ and $\left( {\partial {f_{\mathrm{c}}}/\partial z} \right)/m$ at all heterostructure boundaries.
%%%%%%%%%%%%%%%%%%%%%%%%%%%%%%%%% FIGURE 1 %%%%%%%%%%%%%%%%%%%%%%%%%%
\begin{figure}[h!]
	\centering
		\includegraphics[width=80mm]{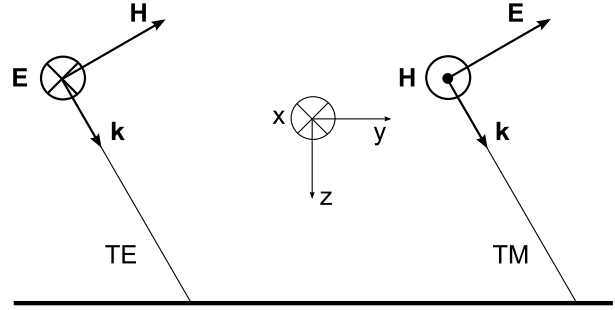}
	\caption{\label{Fig1} Geometry of a transverse electric (TE) wave and a transverse magnetic (TM) wave.}
\end{figure}
%%%%%%%%%%%%%%%%%%%%%%%%%%%%%%%%%%%%%%%%%%%%%%%%%%%%%%%%%%%%%%%%%%%%%

Due to the vector nature of electromagnetic fields, the propagation of electromagnetic plane waves in an isotropic medium, unlike ballistic electrons, depends on polarization. For the purposes of this paper, we define (see Fig. \ref{Fig1}) the TE wave as that characterized by ${k_y},{k_z},{E_x},{H_y},{H_z},\frac{\partial }{{\partial x}} \to 0,\frac{\partial }{{\partial y}} \to {\mathrm{i}}{k_y}$ and the TM wave as that characterized by ${k_y},{k_z},{H_x},{E_y},{E_z},\frac{\partial }{{\partial x}} \to 0,\frac{\partial }{{\partial y}} \to {\mathrm{i}}{k_y}$. The propagation of TE and TM waves in an isotropic material is described by Maxwell's equations, which can be written as 
%%%%%%%%%%%%%%%%%%%%%%%%%%%%%%%%% Equation 4 %%%%%%%%%%%%%%%%%%%%%%%%%
\begin{eqnarray}
	\label{eq4}
		\frac{\partial }{{\partial z}}\left[ {\begin{array}{*{20}{c}}
{{E_x}}\\
{\frac{{ - {\mathrm{i}}}}{{\omega \mu }}\frac{{\partial {E_x}}}{{\partial z}}}
\end{array}} \right] = \left[ {\begin{array}{*{20}{c}}
0&{{\mathrm{i}}\omega \mu }\\
{{\mathrm{i}}\omega \varepsilon \left( {1 - \frac{{k_y^2}}{{{{\omega ^2}\varepsilon \mu}}}} \right)}&0
\end{array}} \right]\left[ {\begin{array}{*{20}{c}}
{{E_x}}\\
{\frac{{ - {\mathrm{i}}}}{{\omega \mu }}\frac{{\partial {E_x}}}{{\partial z}}}
\end{array}} \right]
\end{eqnarray}
%%%%%%%%%%%%%%%%%%%%%%%%%%%%%%%%%%%%%%%%%%%%%%%%%%%%%%%%%%%%%%%%%%%%%%
for TE wave and 
%%%%%%%%%%%%%%%%%%%%%%%%%%%%%%%%% Equation 5 %%%%%%%%%%%%%%%%%%%%%%%%%
\begin{eqnarray}
	\label{eq5}
		\frac{\partial }{{\partial z}}\left[ {\begin{array}{*{20}{c}}
{{H_x}}\\
{\frac{{ - {\mathrm{i}}}}{{\omega \varepsilon }}\frac{{\partial {H_x}}}{{\partial z}}}
\end{array}} \right] = \left[ {\begin{array}{*{20}{c}}
0&{{\mathrm{i}}\omega \varepsilon }\\
{{\mathrm{i}}\omega \mu \left( {1 - \frac{{k_y^2}}{{{{\omega ^2}\varepsilon \mu}}}} \right)}&0
\end{array}} \right]\left[ {\begin{array}{*{20}{c}}
{{H_x}}\\
{\frac{{ - {\mathrm{i}}}}{{\omega \varepsilon }}\frac{{\partial {H_x}}}{{\partial z}}}
\end{array}} \right]
\end{eqnarray}
%%%%%%%%%%%%%%%%%%%%%%%%%%%%%%%%%%%%%%%%%%%%%%%%%%%%%%%%%%%%%%%%%%%%%%
for TM wave. The boundary conditions in this case demand the continuity of $E_x$, $H_x$, $\left( {\partial {E_x}/\partial z} \right)/\mu$ and $\left( {\partial {H_x}/\partial z} \right)/\varepsilon$.

By mutual comparison of (\ref{eq3}) with (\ref{eq4}) and (\ref{eq5}), the analogy between electron waves and electromagnetic plane waves can be written as
%%%%%%%%%%%%%%%%%%%%%%%%%%%%%%% Table 1 %%%%%%%%%%%%%%%%%%%%%%%%%%%%%%
\begin{table}[ht]
	\begin{center}
		\begin{tabular}{|c|c|c|} 
		\hline 
		\textbf{Electron wave} 		& \textbf{TE plane wave} 	& \textbf{TM plane wave} \\ \hline 
		$f_\mathrm{c}$ 								 		& $E_x$ 									& $H_x$ \\ \hline 
		$m$			 							 		& $\mu$				 						& $\epsilon$ \\ \hline 
		$\Delta E=2\left( {E - E_{\Gamma_6}} \right)$ & $\epsilon$ 	& $\mu$ \\ \hline 
		$k^2 ={m \Delta E/{\hbar ^2}}$ & ${\omega ^2}\varepsilon \mu$ & 				${\omega ^2}\varepsilon \mu$ \\ \hline
		\end{tabular}
		\caption{\label{tab1} The analogy between electron and plane electromagnetic waves.}
	\end{center}
\end{table}
%%%%%%%%%%%%%%%%%%%%%%%%%%%%%%%%%%%%%%%%%%%%%%%%%%%%%%%%%%%%%%%%%%%%%%

%---------------------------------------------------------------------------------------------------
%---------------------------------------------------------------------------------------------------
\section{A Perfect Lens for Electron}
The perfect lens for electromagnetic waves is defined \cite{Pendry-2000} as an isotropic slab of thickness $d_\mathrm{lens}$ with material constants $\varepsilon_\mathrm{in}$ and $\mu_\mathrm{in}$ surrounded by another isotropic material with parameters $\varepsilon_\mathrm{out}$ and $\mu_\mathrm{out}$. If the materials are chosen \cite{Pendry-2000} so that ${\varepsilon _{\mathrm{out}}} =  - {\varepsilon _{\mathrm{in}}}$ and ${\mu _{\mathrm{out}}} =  - {\mu _{\mathrm{in}}}$ the slab behaves as a perfect lens, which transfers all plane waves, including all evanescent harmonics, from the source plane at a distance $d_{\mathrm{source}}$ in front of the lens, to the image plane, which is situated at a distance $d_{\mathrm{image}}$ behind the lens, provided that the distances are chosen such that $k_z^{{\mathrm{out}}}{d_{{\mathrm{source}}}} + k_z^{{\mathrm{out}}}{d_{{\mathrm{image}}}} = k_z^{{\mathrm{in}}}{d_{{\mathrm{lens}}}}$.

Comparing the above mentioned conditions and the analogies from Tab. \ref{tab1}, it can be seen that the quantum analogy of the perfect lens is rather unlikely to exist. This is due to the fact that the mass $m$ and $\Delta E$ are mutually bound by the material band structure. Therefore, these variables cannot be tuned separately, as is the case in electromagnetic metamaterials, in which $\varepsilon$, $\mu$ are usually connected to distinct elements. Fortunately, as shown in \cite{Pendry-2000}, if one is working in quasistatic conditions (${k_{y}} \gg {\omega ^2}\varepsilon \mu$, i.e. in the near field) and is thus not interested in the propagative harmonics, then total transmission of the evanescent spectrum can be achieved by $\varepsilon_{\mathrm{in}}=\mathrm{any}$, $\mu_{\mathrm{in}}=-\mu_{\mathrm{out}}$ for TE waves and $\mu_{\mathrm{in}}=\mathrm{any}$, $\varepsilon_{\mathrm{in}}=-\varepsilon_{\mathrm{out}}$ for TM waves. This system is known as the "poor man's lens".
For an electron lens, this implies the use of a slab of $m<0$ and $\Delta E = \mathrm{any}$. The quasistatic regime in this case is given by ${k_y} \gg k = \sqrt{m \Delta E/{\hbar ^2}} $. In order to put the above--stated conditions on solid grounds, the transmission and reflection coefficient of the lens is written as
%%%%%%%%%%%%%%%%%%%%%%%%%%%%%%%%% Equation 6 %%%%%%%%%%%%%%%%%%%%%%%%%
\begin{eqnarray}
	\label{eq6x}
	\begin{array}{l}
T = \frac{{4{Y_{{\rm{in}}}}{Y_{{\rm{out}}}}{{\rm{e}}^{{\rm{i}}k_z^{{\rm{out}}}{d_{{\rm{source}}}}}}{{\rm{e}}^{{\rm{i}}k_z^{{\rm{out}}}{d_{{\rm{image}}}}}}}}{{{{\left( {{Y_{{\rm{in}}}} + {Y_{{\rm{out}}}}} \right)}^2}{{\rm{e}}^{ - {\rm{i}}k_z^{{\rm{in}}}{d_{{\rm{lens}}}}}} - {{\left( {{Y_{{\rm{in}}}} - {Y_{{\rm{out}}}}} \right)}^2}{{\rm{e}}^{{\rm{i}}k_z^{{\rm{in}}}{d_{{\rm{lens}}}}}}}}\\
\\
R = \frac{{\left( {Y_{{\rm{in}}}^2 - Y_{{\rm{out}}}^2} \right)\left( {{{\rm{e}}^{{\rm{i}}k_z^{{\rm{in}}}{d_{{\rm{lens}}}}}} - {{\rm{e}}^{ - {\rm{i}}k_z^{{\rm{in}}}{d_{{\rm{lens}}}}}}} \right){{\rm{e}}^{2{\rm{i}}k_z^{{\rm{out}}}{d_{{\rm{source}}}}}}}}{{{{\left( {{Y_{{\rm{in}}}} + {Y_{{\rm{out}}}}} \right)}^2}{{\rm{e}}^{ - {\rm{i}}k_z^{{\rm{in}}}{d_{{\rm{lens}}}}}} - {{\left( {{Y_{{\rm{in}}}} - {Y_{{\rm{out}}}}} \right)}^2}{{\rm{e}}^{{\rm{i}}k_z^{{\rm{in}}}{d_{{\rm{lens}}}}}}}},
\end{array}
\end{eqnarray}
%%%%%%%%%%%%%%%%%%%%%%%%%%%%%%%%%%%%%%%%%%%%%%%%%%%%%%%%%%%%%%%%%%%%%%%%%
where $k_z^2 = {k^2} - k_y^2$ and $Y = {k_z}/m$. Clearly, in the case of the perfect lens (${\Delta E_{\mathrm{out}}} =  - {\Delta E_{\mathrm{in}}}$, ${m_{\mathrm{out}}} =  - {m_{\mathrm{in}}}$, ${d_{{\mathrm{source}}}} + {d_{{\mathrm{image}}}} = {d_{{\mathrm{lens}}}}$) there is $k_z^{{\rm{in}}} = k_z^{{\rm{out}}}$, ${Y_{{\rm{in}}}} = -{Y_{{\rm{out}}}}$ and thus $T = 1$ and $R=0$, irrespective of $k_y$. The case of the poor man's lens is a little more tricky. Assume that the waves are generated by a source at $z<0$. For $k_y > k$ the physicality of the solution (wave decays in the direction of propagation) demands $k_z = \mathrm{i} \alpha$ with $\alpha >0$. Assume further that ${m_{\mathrm{out}}} =  - {m_{\mathrm{in}}}$ and ${d_{{\mathrm{source}}}} + {d_{{\mathrm{image}}}} = {d_{{\mathrm{lens}}}}$. Two limits of (\ref{eq6x}) are then of interest. First, for $k_y \gg k$ but $\left| {{k_z}{d_{{\rm{lens}}}}} \right| \ll 1$ there is $k_z^{{\rm{in}}} \to k_z^{{\rm{out}}}$, ${Y_{{\rm{in}}}} \to  - {Y_{{\rm{out}}}}$ and $T \to 1$, $R \to 0$. The second limit is ${k_y} \to \infty$ with ${\Delta E_{\mathrm{out}}} \neq  - {\Delta E_{\mathrm{in}}}$. In this case, (\ref{eq6x}) goes to 
%%%%%%%%%%%%%%%%%%%%%%%%%%%%%%%%% Equation 7 %%%%%%%%%%%%%%%%%%%%%%%%%
\begin{eqnarray}
	\label{eq7x}
\begin{array}{l}
T \approx  - {\left( {\frac{{4{\hbar ^2}k_y^2}}{{{m_{{\mathrm{in}}}}\left( {\Delta {E_{{\mathrm{in}}}} + \Delta {E_{{\mathrm{out}}}}} \right)}}} \right)^2}{{\mathrm{e}}^{ - 2{k_y}{d_{{\mathrm{lens}}}}}} \approx 0\\
\\
R \approx \frac{{4{\hbar ^2}k_y^2}}{{{m_{{\mathrm{in}}}}\left( {\Delta {E_{{\mathrm{in}}}} + \Delta {E_{{\mathrm{out}}}}} \right)}}{{\mathrm{e}}^{ - 2{k_y}{d_{{\mathrm{source}}}}}} \approx 0.
\end{array}
\end{eqnarray}
%%%%%%%%%%%%%%%%%%%%%%%%%%%%%%%%%%%%%%%%%%%%%%%%%%%%%%%%%%%%%%%%%%%%%%%%%
In the case of the poor man's lens, there thus clearly exist only a finite band of $k_y$ for which $T \approx 1$. The bandwidth grows as the lens gets thinner and also as $k$ gets smaller.
%---------------------------------------------------------------------------------------------------
%---------------------------------------------------------------------------------------------------
\section{Application of the Theory}
The Kane's model used for the above reasoning assumes that the bulk band edge eigenfunctions are the same, or very similar, throughout the heterostructure \cite{Bastard}. This condition can best be met by the use of the same material inside and outside of the lens. In order to meet the criterion of ${m_{{\mathrm{in}}}} =  - {m_{{\mathrm{out}}}}$ in such a case, a static energy shift imposed by a voltage gate can be used. Since the presence of an electric field is not taken into account in the model, the static shift must be kept small to obtain meaningful results. For this reason, the most suitable material for creating the lens is a material which, under a small static shift, will change the mass from a positive value to a negative value at given energy. In addition, however, the outer material needs to be propagative at $k_y=0$. Consulting the definition of effective mass (\ref{eq2}), the only way to achieve these conditions seems to be a semiconductor with a so--called inverted band structure, i.e. the $\Gamma_6$ band edge positioned below the $\Gamma_8$ band edge. As will be justified below, a good candidate for such a material is HgTe which satisfies all the above criteria.

\begin{figure}
	\centering
			\includegraphics[width=160mm]{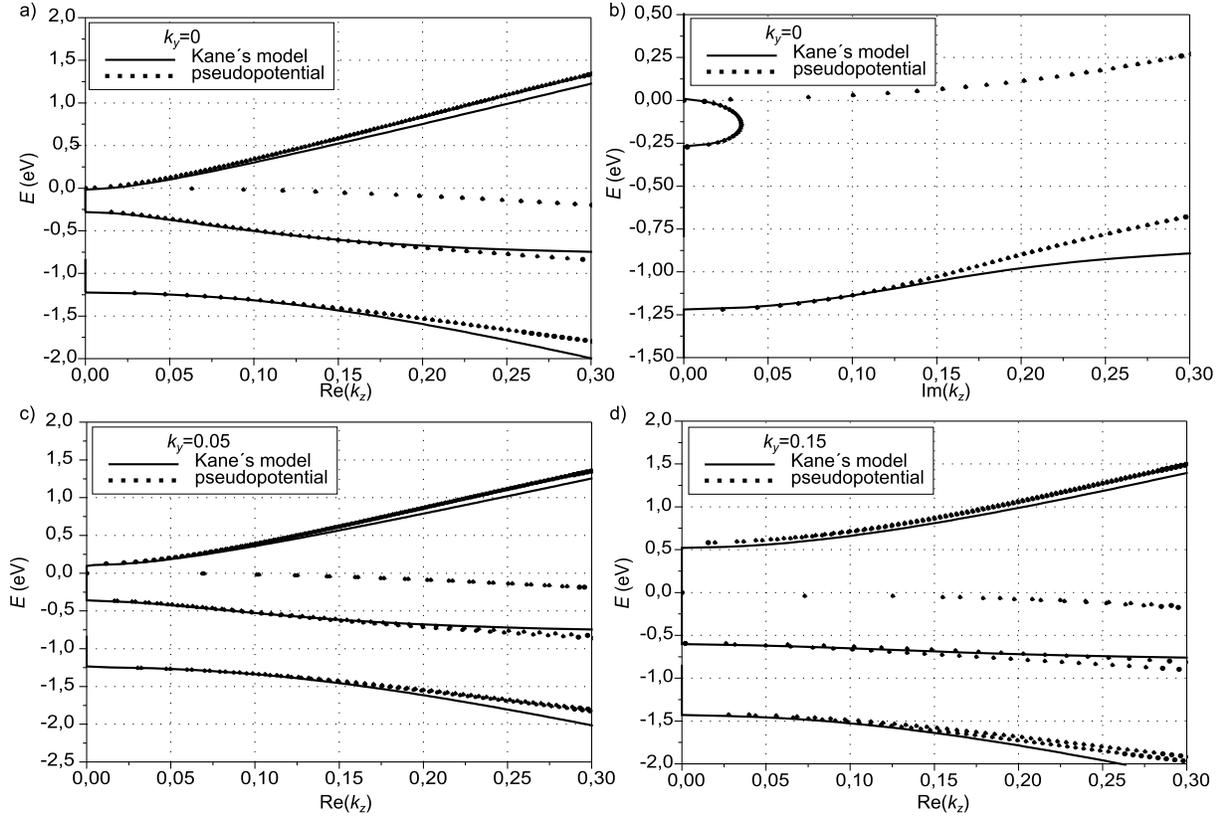}
	\caption{\label{fig2x} A comparison of the HgTe band structure computed by the Kane's model (solid) and by the local pseudopotential method (dotted). The wavenumber is normalized to $2\pi /a$, where $a$ is the lattice constant.}
\end{figure}

In order to make quantitative proposals, the Kane's model needs to be supplied with parameter $2m_{\mathrm{e}}P^2$, and the band edge energies $E_{\Gamma_6}$, $E_{\Gamma_7}$ and $E_{\Gamma_8}$. For this purpose, the HgTe band structure was calculated the with local pseudopotential method (LPM) \cite{Bloom-1970}, \cite{Potz-1981}. The pseudopotential form factors, structure factors and spin-orbit parameters from \cite{Mecabih-2000} were used. The calculation used 137 reciprocal vectors, which were sufficient to reach convergence. A comparison of the Kane's model and LPM is shown in Fig. \ref{fig2x} for several cases. As can be seen, the two methods are very similar for small wavenumbers. The differences are caused by approximations underlining the Kane's model, mainly the small number of bands that are used (4-band model), see Fig. \ref{fig2x}a and Fig. \ref{fig2x}b, and omission of the spin-states splitting at $k_y \ne 0$. This last limitation is well visible at higher $k_y$, where the spin--states calculated by LPM are clearly non-degenerated, unlike in the Kane's model, see Fig. \ref{fig2x}c and Fig. \ref{fig2x}d. Note also that the Kane's model completely ignores heavy holes. Their coupling to light states is however very weak and their possible addition has no practical influence.  For $\left\| {{\bf{k}}a/\left( {2\pi } \right)} \right\| < 0.3$, with $a$ being the size of the unit cell, the Kane's model can however be denoted as a good approximation, and will be used in subsequent calculations. 

%%%%%%%%%%%%%%%%%%%%%%%%%%%%%%%%%%%%%%%%% Figure 2 %%%%%%%%%%%%%%%%%%%%%%%%%
\begin{figure}
	\centering
		\includegraphics[width=160mm]{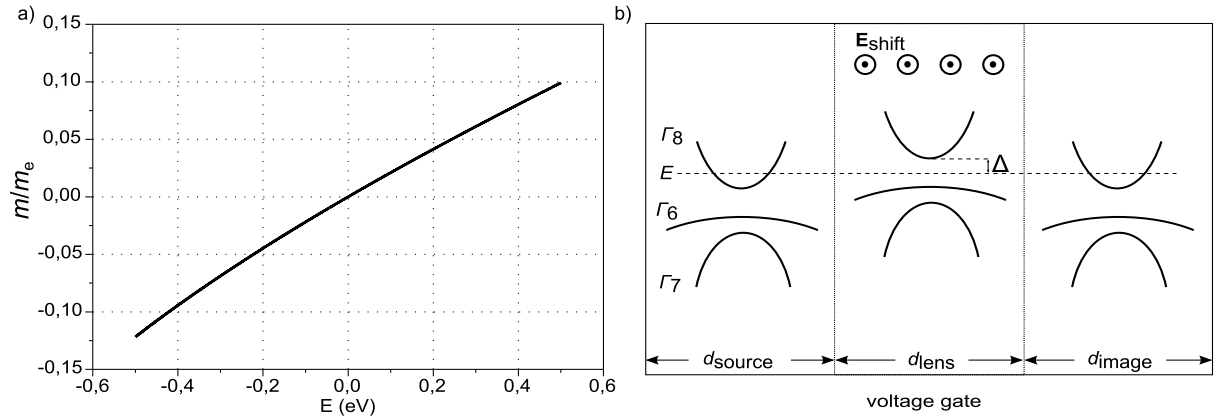}
	\caption{\label{fig3x} (a) Normalized mass as a function of energy in the vicinity of $E_{\Gamma_8}$.  (b) Sketch of the poor man's lens. The lens structure is created by an applied voltage producing a homogeneous electric field that shifts the band structure by $2\Delta$. The electrodes supplying the energy shift cover the whole surface.}
\end{figure}

%%%%%%%%%%%%%%%%%%%%%%%%%%%%%%%%%%%%%%%%%%%%%%%%%%%%%%%%%%%%%%%%%%%%%%%%%%%%
Figure \ref{fig3x}a shows the effective mass (\ref{eq2}) of HgTe in the vicinity of $E_{\Gamma_8}$. The mass clearly changes sign at $E_{\Gamma_8}$, being practically a linear function of energy in the plotted range. 
%%%%%%%%%%%%%%%%%%%%%%%%%%%%%%%%%%%%%%%%% Figure 3 %%%%%%%%%%%%%%%%%%%%%%%%%
%%%%%%%%%%%%%%%%%%%%%%%%%%%%%%%%%%%%%%%%%%%%%%%%%%%%%%%%%%%%%%%%%%%%%%%%%%%%
Based on this observation, the structure of the quantum poor man's lens is depicted in Fig.\ref{fig3x}b. The energy shift in the middle part of the heterostructure induced by the voltage gate is set so that at given energy $E$ the mass just changes sign when traversing the interface. Using the linear dependence of the effective mass shown in Fig. \ref{fig3x}a, this means that $\Delta =  E - E_{\Gamma_8} $. At this point it is also important to mention that in this way the absolute value of the mass can be made as small as desired and $k =\sqrt{m \Delta E/{\hbar ^2}}$ with it, enlarging the operation bandwidth (in $k_y$) of the lens, see the discussion in Sec. 3.

\begin{figure}
	\centering
		\includegraphics[width=160mm]{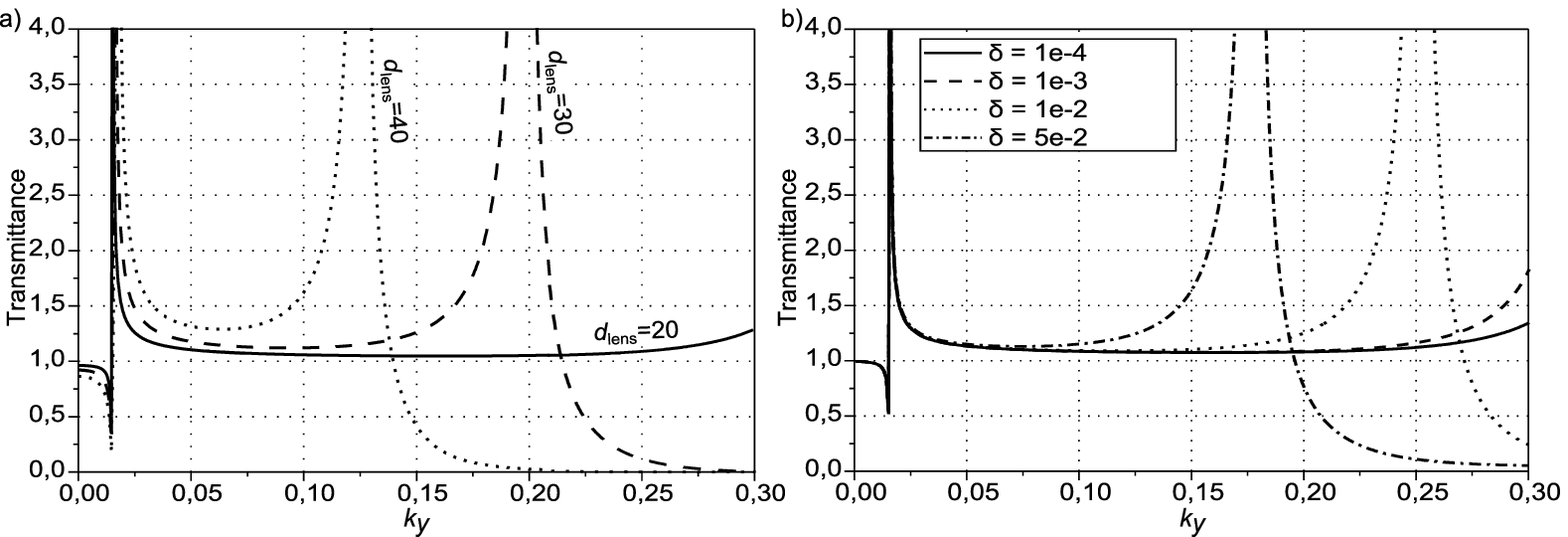}
	\caption{\label{fig4x} Transmittance as a function of ${k_{\mathrm{y}}}$. The wavenumber is normalized to $2\pi /a$ and distances to $a/\left( {2\pi } \right)$, where $a$ is the lattice constant. Panel (a) shows the dependence on the lens thickness $d_{\mathrm{lens}}$, keeping ${m_{{\mathrm{out}}}} = - {m_{{\mathrm{in}}}}$. Panel (b) shows the dependence on $\delta = 2\left( {{m_{{\mathrm{out}}}} + {m_{{\mathrm{in}}}}} \right)/\left( {{m_{{\mathrm{out}}}} - {m_{{\mathrm{in}}}}} \right)$, keeping $d_{\mathrm{lens}} = 20$.}
\end{figure}

The transmittance through the lens heterostructure is depicted in Fig.\ref{fig4x}a for several values of ${d_{{\mathrm{lens}}}}$ and for $m_{{\mathrm{out}}} = - m_{{\mathrm{in}}}$. 
%%%%%%%%%%%%%%%%%%%%%%%%%%%%%%%%%%%%%%%%% Figure 4 %%%%%%%%%%%%%%%%%%%%%%%%%
%%%%%%%%%%%%%%%%%%%%%%%%%%%%%%%%%%%%%%%%%%%%%%%%%%%%%%%%%%%%%%%%%%%%%%%%%%%%
The transmittance curves clearly show the limited $k_y$ bandwidth discussed in Sec. 3. In fact, based on these curves, the bounds of this band can be estimated, since the useful band evidently lies between the two poles of the transmission coefficient (\ref{eq6x}). The left pole appears at $k_y = k_{{\mathrm{in}}}$. The right pole for ${k_y} \gg {k_{{\mathrm{in}}}},{k_{{\mathrm{out}}}}$ approximately satisfies
%%%%%%%%%%%%%%%%%%%%%%%%%%%%%%%%% Equation 8 %%%%%%%%%%%%%%%%%%%%%%%%%
\begin{eqnarray}
	\label{eq8x}
{d_{{\mathrm{lens}}}} = \frac{1}{{{k_y}}}\ln \left| {\frac{{4{\hbar ^2}k_y^2}}{{{m_{{\mathrm{in}}}}\left( {\Delta {E_{{\mathrm{in}}}} + \Delta {E_{{\mathrm{out}}}}} \right)}}} \right|.
\end{eqnarray}
%%%%%%%%%%%%%%%%%%%%%%%%%%%%%%%%%%%%%%%%%%%%%%%%%%%%%%%%%%%%%%%%%%%%%%%%%
This equation, for given $k_y$, gives the upper bound on the thickness of the lens which would be able to transmit such a harmonic. Finally, the transmittance is also plotted in Fig.\ref{fig4x}b, but now for given thickness ${d_{{\mathrm{lens}}}}$ and for several values of mass deviation $\delta = 2\left( {{m_{{\mathrm{out}}}} + {m_{{\mathrm{in}}}}} \right)/\left( {{m_{{\mathrm{out}}}} - {m_{{\mathrm{in}}}}} \right)$. The curves show that the lens is also sensitive in this respect, and that the shifting voltage will have to be fine--tuned in real designs. 

\section{Conclusions}
The analogy between electron waves and light waves has been studied. It has been shown that an almost exact analogy can be achieved between electromagnetic plane waves and ballistic electrons obliquely incident on a planar boundary. Based on this analogy, a quantum perfect lens has been discussed. It has been shown that constraints related to a realistic electronic band structure almost preclude the existence of a perfect electron lens. However, a quasi-static version of the lens (a poor man's lens) has been shown to be feasible. Using an HgTe semiconductor compound, a particular design has been proposed and numerically investigated. 

\ack{This work has been supported by the Czech Science Foundation under project 13-09086S and P108/11/0894, and by the Czech Technical University in Prague under project OHK3-011/13.}

\section*{References}
%\bibliography{References2013}
%\bibliographystyle{unsrt}

\end{document}